**Si Metasurface Supporting Multiple Quasi-BICs for Degenerate Four-Wave Mixing**

*Gianni Q. Moretti[1,2], Thomas Weber[3], Thomas Possmayer[3], Emiliano Cortés[3],*

*Leonardo de S. Menezes[3,4], Andrea V. Bragas[1,2], Stefan A. Maier[5,6], Andreas Tittl[3*],*

*Gustavo Grinblat[1,2*]*

[1]Universidad de Buenos Aires, Facultad de Ciencias Exactas y Naturales, Departamento de Física, 1428 Buenos Aires, Argentina.

[2]CONICET - Universidad de Buenos Aires, Instituto de Física de Buenos Aires (IFIBA). 1428 Buenos Aires, Argentina.

[3]Chair in Hybrid Nanosystems, Nanoinstitute Munich, Faculty of Physics, Ludwig-Maximilians-Universität München, 80539 München, Germany.

[4]Departamento de Física, Universidade Federal de pernambuco, 50670-901 Recife-PE, Brazil.

[5]School of Physics and Astronomy, Monash University, Clayton Victoria 3800, Australia.

[6]Department of Physics, Imperial College London, London SW7 2AZ, United Kingdom.

*Email: andreas.tittl@physik.uni-muenchen.de; grinblat@df.uba.ar

**Keywords**

**Abstract.** Dielectric metasurfaces supporting quasi-bound states in the continuum (qBICs) enable high field enhancement with narrow-linewidth resonances in the visible and near-infrared ranges. The resonance emerges when distorting the meta-atom's geometry away from a symmetry-protected BIC condition and, usually, a given design can sustain one or two of these states. In this work, we introduce a silicon-on-silica metasurface that simultaneously supports up to four qBIC resonances in the near-infrared region. This is achieved by combining multiple symmetry-breaking distortions on an elliptical cylinder array. By pumping two of these resonances, the nonlinear process of degenerate four-wave mixing is experimentally realized. By comparing the nonlinear response with that of an unpatterned silicon film, the near-field enhancement inside the nanostructured dielectric is revealed. The presented results demonstrate independent geometric control of multiple qBICs and their interaction trough wave mixing processes, opening new research pathways in nanophotonics, with potential applications in information multiplexing, multi-wavelength sensing and nonlinear imaging.

**1. Introduction**

In recent years, metasurfaces supporting resonances originating from bound states in the continuum (BICs) have enabled strong enhancement of electromagnetic fields with high quality factors (Q = $\nu/\Delta\nu$, with $\nu$ the resonance frequency and $\Delta\nu$ its linewidth) in the visible and near-infrared spectral regions. BICs, by nature, are perfectly confined states that cannot couple to the radiation continuum. However, by distorting the meta-atom's shape or orientation of a metasurface supporting a symmetry-protected BIC[1–3], the dark state can be turned into an accessible high-Q quasi-BIC (qBIC). Among many different design choices of unit cells, such as rods[4–6], disks[7, 8], blocks[9, 10], and rings[11, 12], that have allowed quality factors as high as $10^4$, the dimer of tilted elliptical cylinders has shown great promise thanks to its fabrication robustness[13]. Initially devised for realizing optomechanically induced chirality[14], it was later adapted for surface enhanced molecular detection in the mid-infrared region[15] and efficient second harmonic generation with continuous wave laser sources[16].

Devices incorporating nonlinear metasurfaces could be the next step in the emerging industry of flat-optics[17–21], where some uses like nonlinear beam steering[22], directional emission switching[23], nonlinear lensing and imaging[24], and nonlinear holography[25] have already been demonstrated. The applications are, however, restricted by the inherent low efficiency of nonlinear effects in nano-sized media, as the subwavelength interaction volume prevents exploiting phase-matching processes[26]. Dielectric metasurfaces that support qBIC resonances could overcome this problem as they allow for high enhancement of the near fields and exhibit low absorption and high nonlinear susceptibilities in the visible and near-infrared ranges, making them a very interesting platform to further develop the area. QBIC-supporting metasurfaces have demonstrated their capabilities for second-[16, 27] and third-[4, 9, 28, 29] harmonic generation, while more complex nonlinear processes involving multiple frequencies, like four-wave mixing (FWM), have been explored only with Mie-like modes[30, 31], or combining qBIC with Mie resonances[32], but not by mixing different qBICs.

Typically, a single qBIC resonance is present for a given metasurface design[4–10]. Having various of these states could potentially improve the metasurface performance and expand the scope of planar opto-devices. Recently, some theoretical investigations have predicted multiple qBICs in a dielectric metasurface [33–35]. Experimentally, the linear excitation of two qBICs has been demonstrated[36], while stacked metasurfaces have been needed to excite more than two states [37]. In this work, we design and fabricate a metasurface of amorphous-Si (a-Si) elliptical cylinder dimers on silica and exploit two geometric distortions, one involving the tilt angle α of the meta-atoms, and the other the change of intra-dimer separation, d×P/2, with d a dimensionless parameter and P the array period (see Figure 1). We show that two qBIC resonances can be excited for each of these distortions, with all four states present when both asymmetries are introduced. We further study the third order process of degenerate FWM (DFWM), which involves two incident light sources, to exploit the different qBIC resonances of our metasurface, an effect that has so far only been addressed numerically[35, 38]. DFWM combines two photons of frequency $\omega_1$ with one photon of frequency $\omega_2$ to create a photon of frequency $\omega_s = 2\omega_1 - \omega_2$. The experimental results are validated by linear and nonlinear numerical simulations, showing very good agreement.

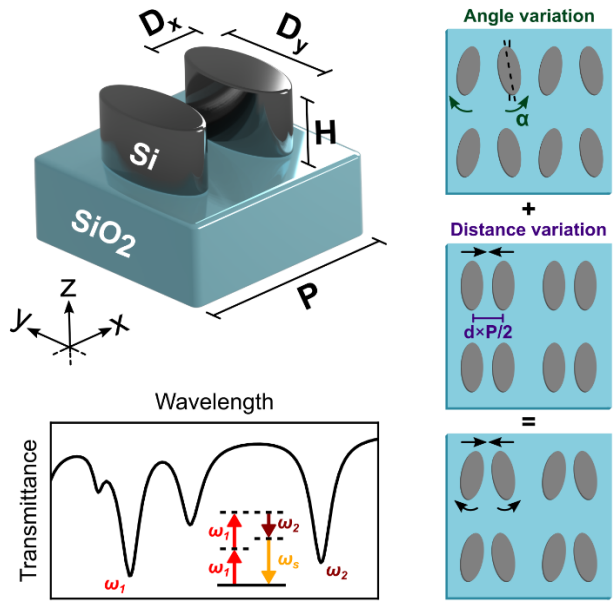

*Figure 1: Metasurface design. The unit cell is composed of elliptical cylinder dimers of fixed height (H), period (P), and diameters ($D_x$, $D_y$). The tilt angle α and the intra-cell distance d×P/2 are varied. The design is made of amorphous silicon (a-Si) on a silica substrate. Two qBIC resonances are exploited for the process of degenerate four-wave mixing.*

## 2. Results

Figure 1 illustrates the unit cell design of the studied a-Si metasurface on silica, fabricated by electron beam lithography (details on the fabrication process can be found in the Methods section). The height (H) is 160 nm, the short ($D_x$) and long ($D_y$) diameters of the elliptical cylinders are 130 and 290 nm, respectively, and the periodicity (P) is 485 nm. These dimensions were chosen to generate resonances in the near-infrared region. The tilt angle (α) and the intra-cell meta-atom separation parameter (d) vary in the ranges 0 to 15° and 0.7 to 1, respectively. When d = 1, the distance between intra- and inter-cell meta-atoms is the same. Representative scanning electron microscope (SEM) images of the metasurfaces can be seen in Figure S1 of the Supplementary Materials.

Figure 2A displays the experimental and simulated transmittance spectra of the metasurfaces by varying d at α = 0° (see Methods for measurement and simulation specifics). The topmost curve (d = 1) shows a broad Mie-like resonance below 700 nm, arising from a magnetic dipole contribution of the individual meta-atom, and a high flat transmission level above 720 nm, within the transparency window of the dielectric. When d is decreased, two narrow resonances appear around 740-750 nm due to the introduced asymmetry, which comes from the difference between the center-to-center distance of meta-atoms within a unit cell and that across neighboring unit cells[39, 40]. A similar scenario develops in Figure 2B, which presents the response when changing α at d = 1. As the tilt angle increases, two resonances emerge around 760 and 820 nm, respectively. In all cases, the resonance linewidths increase when further deviating from the symmetric conditions, with quality factors ranging from 50 to 200.

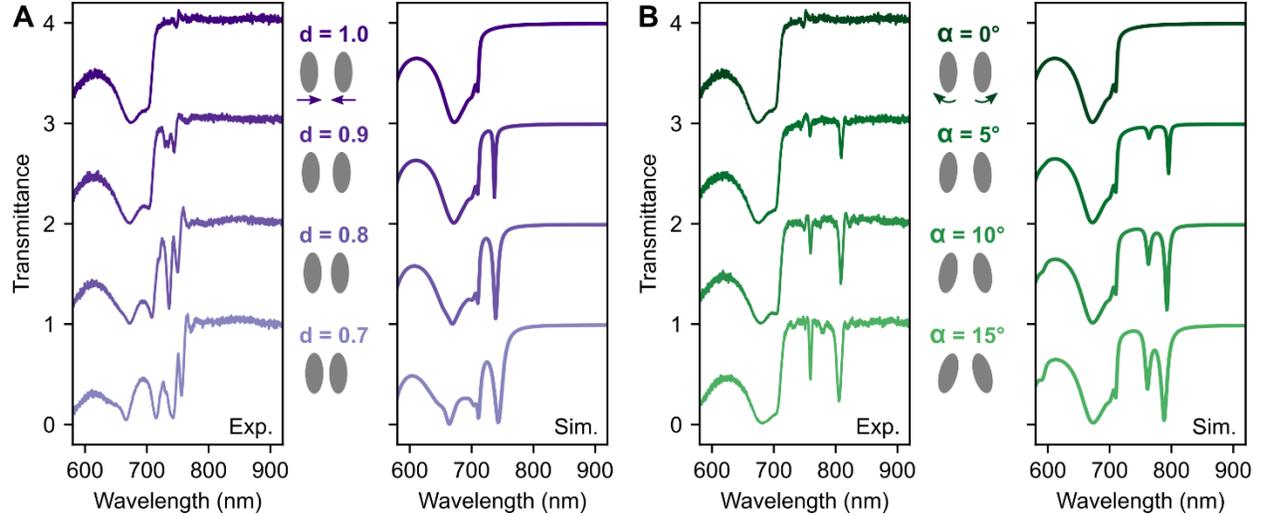

*Figure 2: QBIC resonances for different geometric variations. (A) Measured (left) and simulated (right) transmittance spectra of the metasurface when decreasing d at α = 0° for x-polarized incoming light. A schematic top-view of the dimer meta-molecule is displayed between graphs for each geometric variation. (B) Same as (A) but for increasing α at d = 1. The transmittance spectra are vertically displaced for clarity.*

Next, we demonstrate that the in-plane rotation and lateral displacement of the meta-atoms can be combined to simultaneously generate the two pairs of qBICs, as can be seen in Figure 3A,B by setting d = 0.8 and α = 10°. The average field enhancement inside the dielectric is presented in Figure 3C, showing an enhancement factor around 3 at resonant conditions. To characterize this light confinement, we later performed DFWM experiments by exciting two of the most prominent resonances, at 741 and 816 nm (740 and 800 nm in simulations, respectively), marked with asterisks in the graphs. Their corresponding field distributions at different cross-sections of the unit cell can be seen in Figure 3D. At 800 nm (right column), an electric dipole pointing in the x-direction allows x-polarized light to excite the qBIC, due to the tilt distortion. This state, which has been extensively observed in tilted elliptical cylinders metasurfaces [15, 41, 42], is characterized by an out-of-plane magnetic dipole and an in-plane electric quadrupole that cancel each other out in the far-field when α = 0°; increasing the tilt angle adds an electric dipole component that couples the state to external radiation. As for the other selected qBIC, at 740 nm wavelength (left column in Figure 3D), the excitation of a magnetic dipole in the y-direction, originating from the asymmetry parameter d, creates an out-of-plane electric field circulation, as has been previously observed in arrays of dimers of circular[33, 43] and D-shaped[44] meta-atoms. The two remaining states (at 730 and 760 nm) correspond to the magnetic counterparts of the described resonances. A detailed mapping of the field distributions for all four resonances with their multipolar decompositions can be found in Figure S2 of the Supplementary Materials. In Figure S3 of the Supplementary Materials, we show the transmission response of other possible combinations of d and α. In Figure S4 of the Supplementary Materials, we demonstrate how all resonances can be tuned across a 100 nm window at a fixed height of the metasurface by scaling the diameters of the elliptical cylinders and the array periodicity. The same figure also reveals that a minimum height of the structure is needed to support the multiple modes, consistent with our previous findings[35].

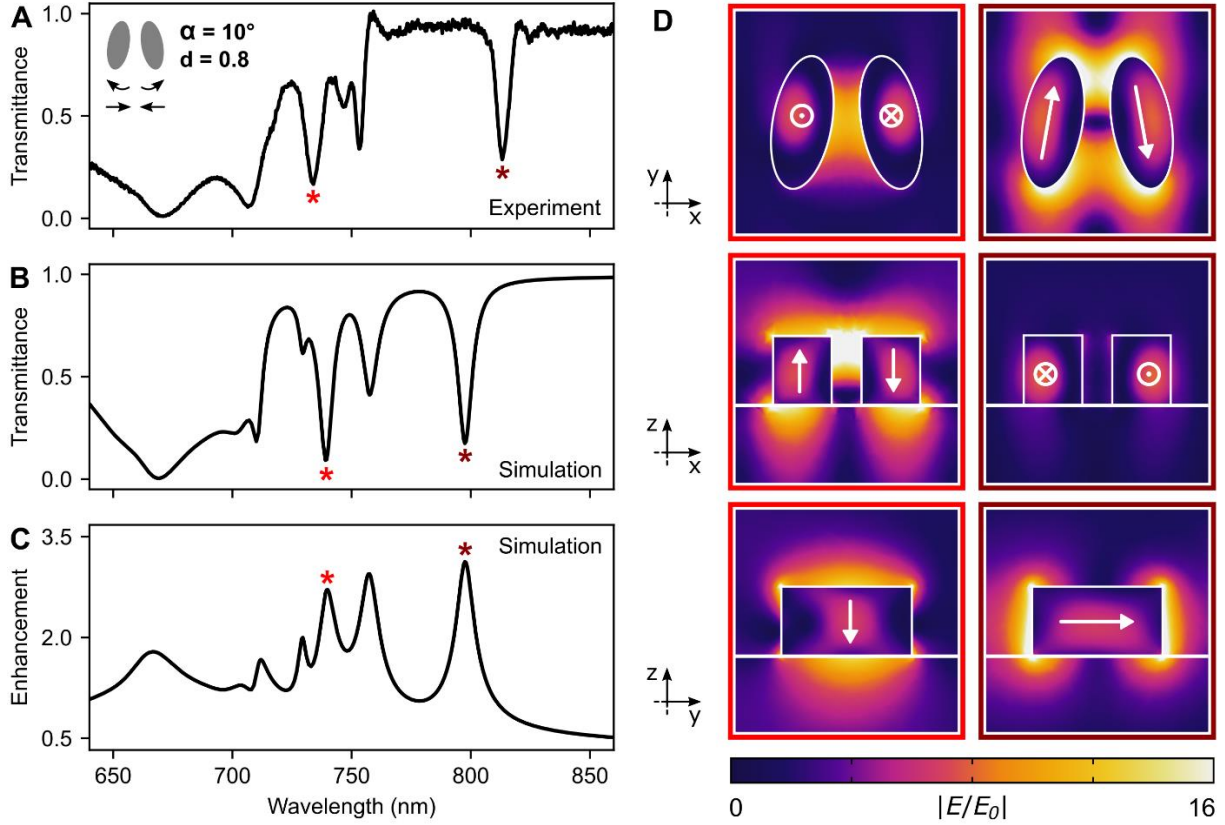

*Figure 3: Combining tilt angle and dimer distance distortions.* (A) Experimental and (B) simulated transmittance spectra of a metasurface with d = 0.8 and α = 10° for x-polarized incoming light. (C) Corresponding average field enhancement inside the dielectric ⟨E/E$_0$⟩. (D) Electric field distributions for different planes of the unit cell at the two resonant conditions highlighted with asterisks in A-C (740 nm, left, and 800 nm, right). These two states were selected for the DFWM experiments.

The DFWM experiments were performed by using two tunable pulsed laser sources, from now on labeled as 'Pump 1', adjustable from 720 to 750 nm, and 'Pump 2', from 800 to 840 nm. Figure 4A shows representative spectra of the pump pulses alongside the signal of the generated nonlinear light at $\omega_s = 2\omega_1 - \omega_2$. By comparing with the response of an unstructured a-Si film, a nonlinear enhancement factor of 600 times is observed, an order of magnitude above that reported for a dimer-hole a-Si metasurface mixing a qBIC and a Mie resonance[32]. To further characterize the nonlinear nature of the process we independently varied the pump power of both lasers ($P_1$ and $P_2$) and monitored the collected nonlinear power ($P_{NL}$), as shown in Figure 4B. As expected, when increasing $P_1$ while keeping $P_2$ constant, $P_{NL}$ shows a quadratic dependence, while a linear relationship is obtained when changing $P_2$ at fixed $P_1$. With this data, we compute a normalized efficiency of the process of $(1.1 \pm 0.2)$ W$^{-2}$%. We attribute this relatively low value to the reduced vectorial overlap between the excited modes. As the nonlinear polarization in the material is maximum when the fields are aligned with one another (see Methods), having mostly in-plane and out-of-plane electric field distributions for the different states, respectively, weakens the nonlinear effect. It is also important to note that these experiments were performed using x-polarized light. Results by varying the polarization are present in Figure S5 of the Supplementary Materials, revealing a further decrease in the signal, as expected from numerical simulations.

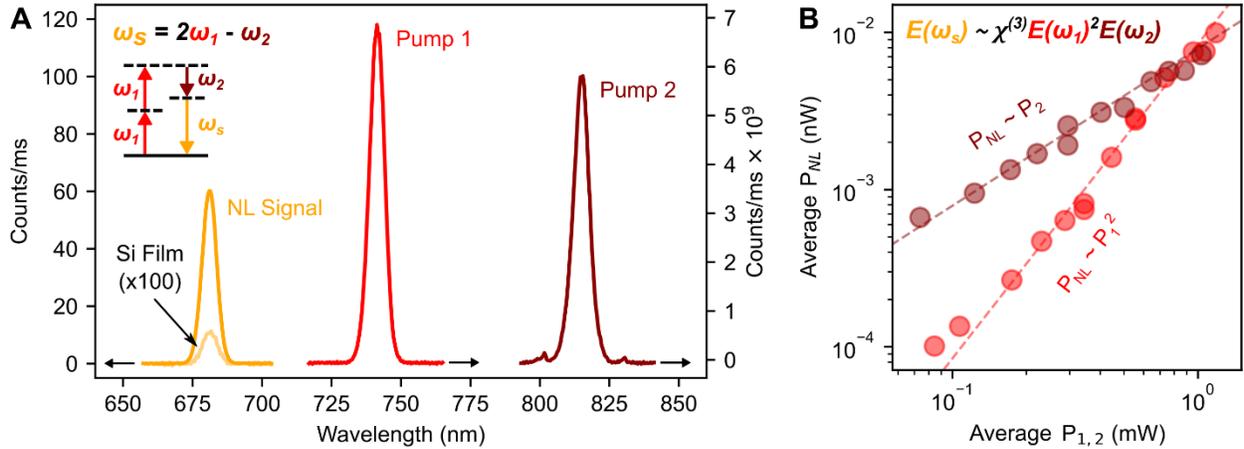

*Figure 4: Degenerate four-wave mixing performance. (A) Pump lasers spectra and nonlinear signal spectrum of the metasurface characterized in Figure 3. The signal of an a-Si film is shown for reference. The nonlinear process diagram is illustrated as an inset. Arrows at the bottom mark the corresponding scale of each spectrum. (B) Dependence of the generated nonlinear light power, $P_{NL}$, on the pump powers $P_1$ and $P_2$. When changing $P_1$, $P_2$ is fixed at 0.76 mW, and a quadratic dependence is obtained. When varying $P_2$ at $P_1 = 0.86$ mW, the relationship is linear.*

To assess the nonlinear signal dependence on the resonant conditions, both pump wavelengths were independently swept. In Figure 5A, on the right side, $\lambda_1$ is kept fixed at 741 nm, and $\lambda_2$ is varied, showing a well-defined peak at the resonant state. On the other hand, fixing $\lambda_2$ = 816 nm and tuning $\lambda_1$ (left side) reveals a less defined peak, which can be described by the contribution of 3 nearby resonances (see Figure 3A), convoluted with the spectral width of the laser, which is around 7-8 nm, same as the studied resonances. Figure 5B exhibits the corresponding nonlinear simulations, showing good agreement with the measurements. Sharper responses are obtained as the calculation assumes monochromatic pumping wavelengths. Given that the measured efficiency was performed in a transmission configuration, the simulated radiation in Figure 5B considers only the cone matching the numerical aperture of our collection objective (see inset), so that the relative height of the peaks can be compared to the measurement.

The presented results show that multiple qBIC resonances can be attained in one metasurface design. We further demonstrate that these states can be tailored to select the enhancement of a specific third-order wave mixing process. This concept could be extended to new mixings based on second-order nonlinearities by using non-centrosymmetric materials, such as GaP, GaAs, and AlGaAs. To improve nonlinear efficiency, fabrication quality as well as spatial and vectorial overlap of the engineered states needs to be optimized.

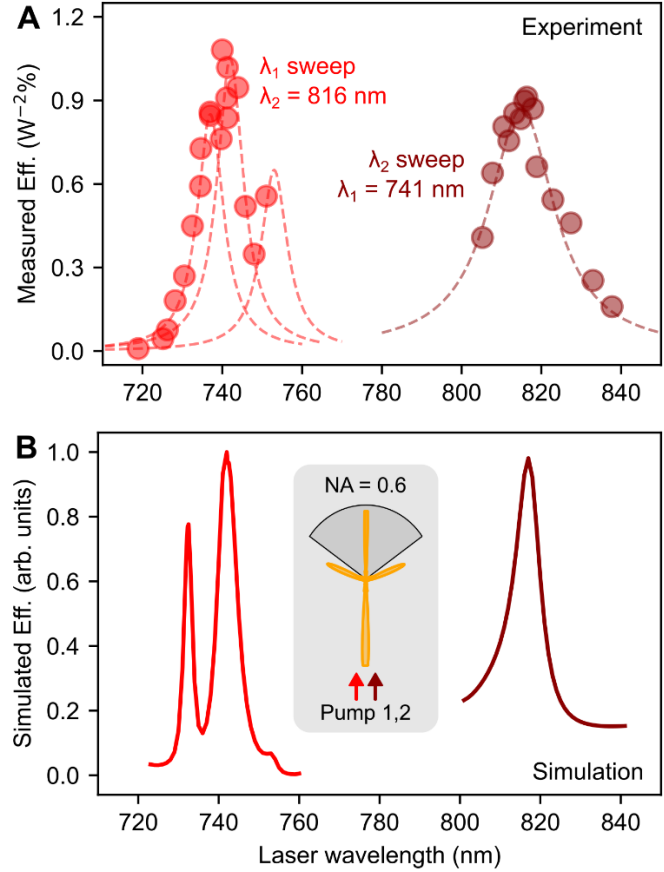

*Figure 5: **Pump wavelengths sweep.** (A) Measured normalized nonlinear efficiency when changing the pump wavelengths λ₁ and λ₂ around the resonant conditions. The dashed lines describe the different resonances contributing to the signal. (B) Simulation of the nonlinear efficiency considering the far-field radiation collected in transmission within an NA = 0.6 to match the experimental conditions (see inset).*

## 3. Conclusion

In summary, we have designed an a-Si on silica metasurface with an elliptical cylinder dimer unit cell that can support two pairs of qBIC resonances, depending on the imposed geometric variations, which are the tilt angle α and the intra-cell distance parameter d. The four qBICs have quality factors ranging from 50 to 200 and can all be present when both asymmetries are introduced. We selected a metasurface with α = 10° and d = 0.8 to study the third order nonlinear process of DFWM by pumping two qBIC resonances. The obtained normalized efficiency is (1.1 ± 0.2) $W^{-2}$% with incident average powers below 1 mW for both laser sources. Our work demonstrates that multiple qBIC resonances can be controlled independently in a metasurface, showing promise for nanoscale wave mixing phenomena. This could enable new research pathways for sensing, nonlinear imaging, and information multiplexing applications of metasurface-based opto-devices.

## 4. Methods

Metasurface fabrication: A 160-nm thick amorphous silicon layer was deposited onto a fused silica substrate by plasma-enhanced chemical vapor deposition (PECVD). The sample was spin-coated with an adhesion layer of SurPass 4000 that was washed away with isopropyl alcohol (IPA) before coating with photoresist ZEP 520A and a highly conducting polymer ESpacer 300Z. The patterning was done via electron-beam lithography (Raith eLINE Plus) with a dosage of 100 µC/$cm^2$. After development for 60 s in 3:1 MIBK:IPA solution, a hard-mask consisting of 30 nm Cr was deposited by electron-beam evaporation before lifting-off the photoresist with a bath of Microposit Remover 1165 for 2 h at 80°C. At this point, the remaining Cr pattern allows to transfer the design by etching the uncovered Si by means of reactive ion etching (RIE). Finally, the Cr layer was removed with Sigma-Aldrich-651826 Standard chromium etchant. The presented SEM images in Figure S1 of the Supplementary Materials were taken with the same e-beam lithography equipment.

Transmittance characterization: A supercontinuum light source from NKT Photonics was used to measure the metasurfaces transmittance spectra. The illumination and the transmitted light collection were done with a 10X, NA = 0.25 and a 60X, NA = 0.70 microscope objectives, respectively.

Nonlinear measurements: Two pulsed sources were provided by a COHERENT Ti:Sapphire laser pumping an OPO module of the same company. 'Pump 1' was selected from the second harmonic of the OPO signal (converting 1000-1600 nm to 500-800 nm) and 'Pump 2' was obtained directly from the Ti:Sapphire laser (700-1000 nm). The pulse width is around 170 fs and the repetition rate 80 MHz. To ensure that both laser pulses arrived at the same time on the sample, a delay line was added on the 'Pump 2' arm. The lasers were focused onto a 10-12 µm spot using a 4X, NA = 0.1 microscope objective. The generated nonlinear light was collected in transmission with a 40X, NA = 0.6 objective. The excitation laser beams were filtered with low pass filters for the nonlinear signal measurements. For more details refer to the schematic in Figure S6 of the Supplementary Materials.

Numerical simulations: Linear and nonlinear calculations were performed using the Wave Optics module of the COMSOL Multiphysics software[45]. The Si structures composing the unit cell were placed on a $SiO_2$/Air interface in a square prism domain geometry with periodic boundary conditions in the four lateral faces and a perfectly matched layer (PML) at the top and bottom faces. The refractive index of the materials was obtained from ellipsometry data; the obtained values for n and k of Si and $SiO_2$ are presented in Figure S7 of the Supplementary Materials. A non-zero extinction coefficient value (k = 0.02) was added to the high-index dielectric to better approximate the experimental results, effectively accounting for non-radiative losses coming from surface roughness.

The nonlinear simulations were performed under a perturbative approximation. First, the linearly excited fields within the nanostructures were computed to define a third-order polarization in the material, which was then used as a source at the DFWM frequency to obtain the nonlinear field. The third-order susceptibility tensor of Si has 21 nonzero components with three independent elements[46]; all were set to $\chi^{(3)} = 2.8\times10^{-18}$ m$^2$/V$^2$[47]. With this, the nonlinear polarization has the form:

$$P(2\omega_1 - \omega_2) = 6\varepsilon_0\chi^{(3)}(E(\omega_1)\cdot E(\omega_2)^*)E(\omega_1) + 3\varepsilon_0\chi^{(3)}(E(\omega_1)\cdot E(\omega_1))E(\omega_2)^*$$

The multipole decompositions shown in Figure S3 of the Supplementary Materials were calculated using the equations found in reference[48]. The near-to-far field transformation used to calibrate the collected signal in Figure 5B was computed using the open-source software package RETOP[49], for which we created an artificial 15 × 15 array using the solution of a single unit cell as the near-field solution.

**Supplementary Material**

This article contains supplementary material.

**Acknowledgments**

This work was partially supported by PICT 2021 IA 363, PICT 2021 GRF-TI-00349, PIP 112 202001 01465, UBACyT Proyecto 20020220200078BA, and UBACyT Proyecto 20020190200296BA. GQM acknowledges 2023 'Ale-Arg' research grant, from the Argentine Ministry of Education and the Deutscher Akademischer Austauschdienst (DAAD, German Academic Exchange Service). AVB acknowledges the Alexander von Humboldt Foundation's generous support. We also acknowledge funding and support from the Deutsche Forschungsgemeinschaft (DFG, German Research Foundation) under

**Conflict of Interest**

The authors declare no conflict of interest.

**Data Availability Statement**

The data supporting this study's findings are available from the corresponding author upon reasonable request.

**References**

[1] C. W. Hsu, B. Zhen, J. Lee, et al., "Observation of trapped light within the radiation continuum," *Nature*, vol. 499, no. 7457, pp. 188–191, 2013, 10.1038/nature12289.

[2] K. Koshelev, S. Lepeshov, M. Liu, A. Bogdanov, and Y. Kivshar, "Asymmetric Metasurfaces with High-Q Resonances Governed by Bound States in the Continuum," *Phys. Rev. Lett.*, vol. 121, no. 19, p. 193903, 2018, 10.1103/PhysRevLett.121.193903.

[3] S. Joseph, S. Pandey, S. Sarkar, and J. Joseph, "Bound states in the continuum in resonant nanostructures: an overview of engineered materials for tailored applications," *Nanophotonics*, vol. 10, no. 17, pp. 4175–4207, 2021, 10.1515/nanoph-2021-0387.

[4] K. Koshelev, Y. Tang, K. Li, D.-Y. Choi, G. Li, and Y. Kivshar, "Nonlinear Metasurfaces Governed by Bound States in the Continuum," *ACS Photonics*, vol. 6, no. 7, pp. 1639–1644, 2019, 10.1021/acsphotonics.9b00700.

[5] M. Gandolfi, A. Tognazzi, D. Rocco, C. De Angelis, and L. Carletti, "Near-unity third-harmonic circular dichroism driven by a quasibound state in the continuum in asymmetric silicon metasurfaces," *Phys. Rev. A*, vol. 104, no. 2, p. 023524, 2021, 10.1103/PhysRevA.104.023524.

[6] L. Kühner, F. J. Wendisch, A. A. Antonov, et al., "Unlocking the out-of-plane dimension for photonic bound states in the continuum to achieve maximum optical chirality," *Light Sci Appl*, vol. 12, no. 1, p. 250, 2023, 10.1038/s41377-023-01295-z.

[7] R. Masoudian Saadabad, L. Huang, and A. E. Miroshnichenko, "Polarization-independent perfect absorber enabled by quasibound states in the continuum," *Phys. Rev. B*, vol. 104, no. 23, p. 235405, 2021, 10.1103/PhysRevB.104.235405.

[8] P. Vaity, H. Gupta, A. Kala, et al., "Polarization-Independent Quasibound States in the Continuum," *Advanced Photonics Research*, vol. 3, no. 2, p. 2100144, 2022, 10.1002/adpr.202100144.

[9] Z. Liu, Y. Xu, Y. Lin, et al., "High-Q Quasibound States in the Continuum for Nonlinear Metasurfaces," *Phys. Rev. Lett.*, vol. 123, no. 25, p. 253901, 2019, 10.1103/PhysRevLett.123.253901.

[10] Y. Zhou, M. Luo, X. Zhao, et al., "Asymmetric tetramer metasurface sensor governed by quasi-bound states in the continuum," *Nanophotonics*, vol. 12, no. 7, pp. 1295–1307, 2023, 10.1515/nanoph-2023-0003.

[11] C. Zhou, X. Qu, S. Xiao, and M. Fan, "Imaging Through a Fano-Resonant Dielectric Metasurface Governed by Quasi-bound States in the Continuum," *Phys. Rev. Applied*, vol. 14, no. 4, p. 044009, 2020, 10.1103/PhysRevApplied.14.044009.

[12] J. Wang, J. Kühne, T. Karamanos, C. Rockstuhl, S. A. Maier, and A. Tittl, "All-Dielectric Crescent Metasurface Sensor Driven by Bound States in the Continuum," *Adv Funct Materials*, vol. 31, no. 46, p. 2104652, 2021, 10.1002/adfm.202104652.

[13] J. Kühne, J. Wang, T. Weber, L. Kühner, S. A. Maier, and A. Tittl, "Fabrication robustness in BIC metasurfaces," *Nanophotonics*, vol. 10, no. 17, pp. 4305–4312, 2021, 10.1515/nanoph-2021-0391.

[14] M. Liu, D. A. Powell, R. Guo, I. V. Shadrivov, and Y. S. Kivshar, "Polarization-Induced Chirality in Metamaterials via Optomechanical Interaction," *Advanced Optical Materials*, vol. 5, no. 16, p. 1600760, 2017, 10.1002/adom.201600760.

[15] A. Tittl, A. Leitis, M. Liu, et al., "Imaging-based molecular barcoding with pixelated dielectric metasurfaces," *Science*, vol. 360, no. 6393, pp. 1105–1109, 2018, 10.1126/science.aas9768.

[16] A. P. Anthur, H. Zhang, R. Paniagua-Dominguez, et al., "Continuous Wave Second Harmonic Generation Enabled by Quasi-Bound-States in the Continuum on Gallium Phosphide Metasurfaces," *Nano Lett.*, vol. 20, no. 12, pp. 8745–8751, 2020, 10.1021/acs.nanolett.0c03601.

[17] M. Khorasaninejad, W. T. Chen, R. C. Devlin, J. Oh, A. Y. Zhu, and F. Capasso, "Metalenses at visible wavelengths: Diffraction-limited focusing and subwavelength resolution imaging," *Science*, vol. 352, no. 6290, pp. 1190–1194, 2016, 10.1126/science.aaf6644.

[18] W. T. Chen, A. Y. Zhu, V. Sanjeev, et al., "A broadband achromatic metalens for focusing and imaging in the visible," *Nature Nanotech*, vol. 13, no. 3, pp. 220–226, 2018, 10.1038/s41565-017-0034-6.

[19] S. Divitt, W. Zhu, C. Zhang, H. J. Lezec, and A. Agrawal, "Ultrafast optical pulse shaping using dielectric metasurfaces," *Science*, vol. 364, no. 6443, pp. 890–894, 2019, 10.1126/science.aav9632.

[20] G. Grinblat, "Nonlinear Dielectric Nanoantennas and Metasurfaces: Frequency Conversion and Wavefront Control," *ACS Photonics*, vol. 8, no. 12, pp. 3406–3432, 2021, 10.1021/acsphotonics.1c01356.

[21] A. I. Kuznetsov, M. L. Brongersma, J. Yao, et al., "Roadmap for Optical Metasurfaces," *ACS Photonics*, p. acsphotonics.3c00457, 2024, 10.1021/acsphotonics.3c00457.

[22] L. Wang, S. Kruk, K. Koshelev, I. Kravchenko, B. Luther-Davies, and Y. Kivshar, "Nonlinear Wavefront Control with All-Dielectric Metasurfaces," *Nano Lett.*, vol. 18, no. 6, pp. 3978–3984, 2018, 10.1021/acs.nanolett.8b01460.

[23] A. Di Francescantonio, A. Zilli, D. Rocco, et al., "All-optical free-space routing of upconverted light by metasurfaces via nonlinear interferometry," *Nat. Nanotechnol.*, 2023, 10.1038/s41565-023-01549-2.

[24] C. Schlickriede, S. S. Kruk, L. Wang, B. Sain, Y. Kivshar, and T. Zentgraf, "Nonlinear Imaging with All-Dielectric Metasurfaces," *Nano Lett.*, vol. 20, no. 6, pp. 4370–4376, 2020, 10.1021/acs.nanolett.0c01105.

[25] Y. Gao, Y. Fan, Y. Wang, W. Yang, Q. Song, and S. Xiao, "Nonlinear Holographic All-Dielectric Metasurfaces," *Nano Lett.*, vol. 18, no. 12, pp. 8054–8061, 2018, 10.1021/acs.nanolett.8b04311.

[26] M. Agio and A. Alù, Eds., *Optical Antennas*, 1st ed. Cambridge University Press, 2013.

[27] L. Kühner, L. Sortino, B. Tilmann, et al., "High-Q Nanophotonics over the Full Visible Spectrum Enabled by Hexagonal Boron Nitride Metasurfaces," *Advanced Materials*, p. 2209688, 2023, 10.1002/adma.202209688.

[28] L. Xu, K. Zangeneh Kamali, L. Huang, et al., "Dynamic Nonlinear Image Tuning through Magnetic Dipole Quasi-BIC Ultrathin Resonators," *Advanced Science*, vol. 6, no. 15, p. 1802119, 2019, 10.1002/advs.201802119.

[29] W. Tong, C. Gong, X. Liu, et al., "Enhanced third harmonic generation in a silicon metasurface using trapped mode," *Opt. Express*, vol. 24, no. 17, p. 19661, 2016, 10.1364/OE.24.019661.

[30] S. Liu, P. P. Vabishchevich, A. Vaskin, et al., "An all-dielectric metasurface as a broadband optical frequency mixer," *Nat Commun*, vol. 9, no. 1, p. 2507, 2018, 10.1038/s41467-018-04944-9.

[31] G. Grinblat, Y. Li, M. P. Nielsen, R. F. Oulton, and S. A. Maier, "Degenerate Four-Wave Mixing in a Multiresonant Germanium Nanodisk," *ACS Photonics*, vol. 4, no. 9, pp. 2144–2149, 2017, 10.1021/acsphotonics.7b00631.

[32] L. Xu, D. A. Smirnova, R. Camacho-Morales, et al., "Enhanced four-wave mixing from multi-resonant silicon dimer-hole membrane metasurfaces," *New J. Phys.*, vol. 24, no. 3, p. 035002, 2022, 10.1088/1367-2630/ac55b2.

[33] Y. He, G. Guo, T. Feng, Y. Xu, and A. E. Miroshnichenko, "Toroidal dipole bound states in the continuum," *Phys. Rev. B*, vol. 98, no. 16, p. 161112, 2018, 10.1103/PhysRevB.98.161112.

[34] J. Tian, Q. Li, P. A. Belov, R. K. Sinha, W. Qian, and M. Qiu, "High-Q All-Dielectric Metasurface: Super and Suppressed Optical Absorption," *ACS Photonics*, vol. 7, no. 6, pp. 1436–1443, 2020, 10.1021/acsphotonics.0c00003.

[35] G. Q. Moretti, E. Cortés, S. A. Maier, A. V. Bragas, and G. Grinblat, "Engineering gallium phosphide nanostructures for efficient nonlinear photonics and enhanced spectroscopies," *Nanophotonics*, vol. 10, no. 17, pp. 4261–4271, 2021, 10.1515/nanoph-2021-0388.

[36] X. Du, L. Xiong, X. Zhao, S. Chen, J. Shi, and G. Li, "Dual-band bound states in the continuum based on hybridization of surface lattice resonances," *Nanophotonics*, vol. 11, no. 21, pp. 4843–4853, 2022, 10.1515/nanoph-2022-0427.

[37] S. C. Malek, A. C. Overvig, A. Alù, and N. Yu, "Multifunctional resonant wavefront-shaping meta-optics based on multilayer and multi-perturbation nonlocal metasurfaces," *Light Sci Appl*, vol. 11, no. 1, p. 246, 2022, 10.1038/s41377-022-00905-6.

[38] T. Liu, M. Qin, F. Wu, and S. Xiao, "High-efficiency optical frequency mixing in an all-dielectric metasurface enabled by multiple bound states in the continuum," *Phys. Rev. B*, vol. 107, no. 7, p. 075441, 2023, 10.1103/PhysRevB.107.075441.

[39] T. Shi, Z.-L. Deng, Q.-A. Tu, Y. Cao, and X. Li, "Displacement-mediated bound states in the continuum in all-dielectric superlattice metasurfaces," *PhotoniX*, vol. 2, no. 1, p. 7, 2021, 10.1186/s43074-021-00029-x.

[40] L. M. Berger, M. Barkey, S. A. Maier, and A. Tittl, "Metallic and All-Dielectric Metasurfaces Sustaining Displacement-Mediated Bound States in the Continuum,"

*Advanced Optical Materials*, vol. 12, no. 5, p. 2301269, 2024, 10.1002/adom.202301269.

[41] M. Liu and D.-Y. Choi, "Extreme Huygens' Metasurfaces Based on Quasi-Bound States in the Continuum," *Nano Lett.*, vol. 18, no. 12, pp. 8062–8069, 2018, 10.1021/acs.nanolett.8b04774.

[42] G. Q. Moretti, A. Tittl, E. Cortés, S. A. Maier, A. V. Bragas, and G. Grinblat, "Introducing a Symmetry-Breaking Coupler into a Dielectric Metasurface Enables Robust High-Q Quasi-BICs," *Advanced Photonics Research*, vol. 3, no. 12, p. 2200111, 2022, 10.1002/adpr.202200111.

[43] S. You, M. Zhou, L. Xu, et al., "Quasi-bound states in the continuum with a stable resonance wavelength in dimer dielectric metasurfaces," *Nanophotonics*, vol. 12, no. 11, pp. 2051–2060, 2023, 10.1515/nanoph-2023-0166.

[44] C. Cui, S. Yuan, X. Qiu, et al., "Light emission driven by magnetic and electric toroidal dipole resonances in a silicon metasurface," *Nanoscale*, vol. 11, no. 30, pp. 14446–14454, 2019, 10.1039/C9NR03172C.

[45] COMSOL Multiphysics® v. 5.4. www.comsol.com. COMSOL AB, Stockholm, Sweden.

[46] R. W. Boyd, *Nonlinear optics, Ch. 1.5*, 3rd ed., San Diego, California, USA: Academic Press, 2008.

[47] N. Bloembergen, W. K. Burns, and M. Matsuoka, "Reflected third harmonic generated by picosecond laser pulses," *Optics Communications*, vol. 1, no. 4, pp. 195–198, 1969, 10.1016/0030-4018(69)90064-9.

[48] E. A. Gurvitz, K. S. Ladutenko, P. A. Dergachev, A. B. Evlyukhin, A. E. Miroshnichenko, and A. S. Shalin, "The High-Order Toroidal Moments and Anapole States in All-Dielectric Photonics," *Laser & Photonics Reviews*, vol. 13, no. 5, p. 1800266, 2019, 10.1002/lpor.201800266.

[49] J. Yang, J.-P. Hugonin, and P. Lalanne, "Near-to-Far Field Transformations for Radiative and Guided Waves," *ACS Photonics*, vol. 3, no. 3, pp. 395–402, 2016, 10.1021/acsphotonics.5b00559.

Supplementary Materials for:

**Si Metasurface Supporting Multiple Quasi-BICs for Degenerate Four-Wave Mixing**

*Gianni Q. Moretti, Thomas Weber, Thomas Possmayer, Emiliano Cortés,*

*Leonardo de S. Menezes, Andrea V. Bragas, Stefan A. Maier, Andreas Tittl[*], Gustavo Grinblat[*]*

*Email: andreas.tittl@physik.uni-muenchen.de; grinblat@df.uba.ar

**Contents**

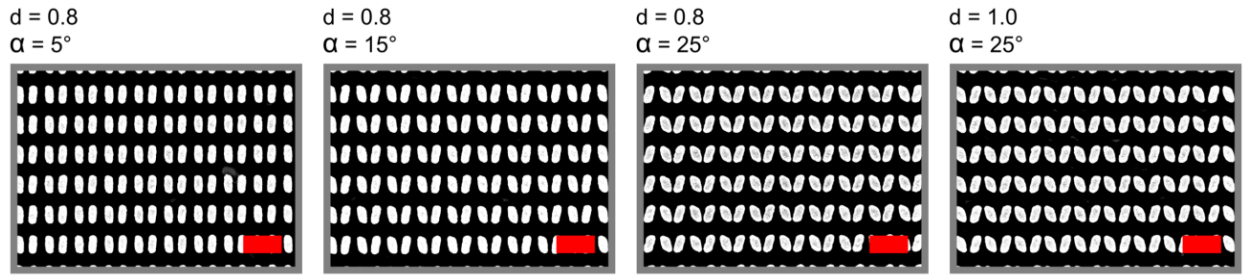

**Figure S1: SEM Images of representative fabricated metasurfaces.** Top-view of some of the studied metasurfaces. From left to right, the first three show increasing angle α at fixed intra-cell distance parameter d. The last two compare the case of non-equidistant vs equidistant meta-atoms (d = 1.0) at fixed α. Red scale bar at the bottom-right corner, 600 nm.

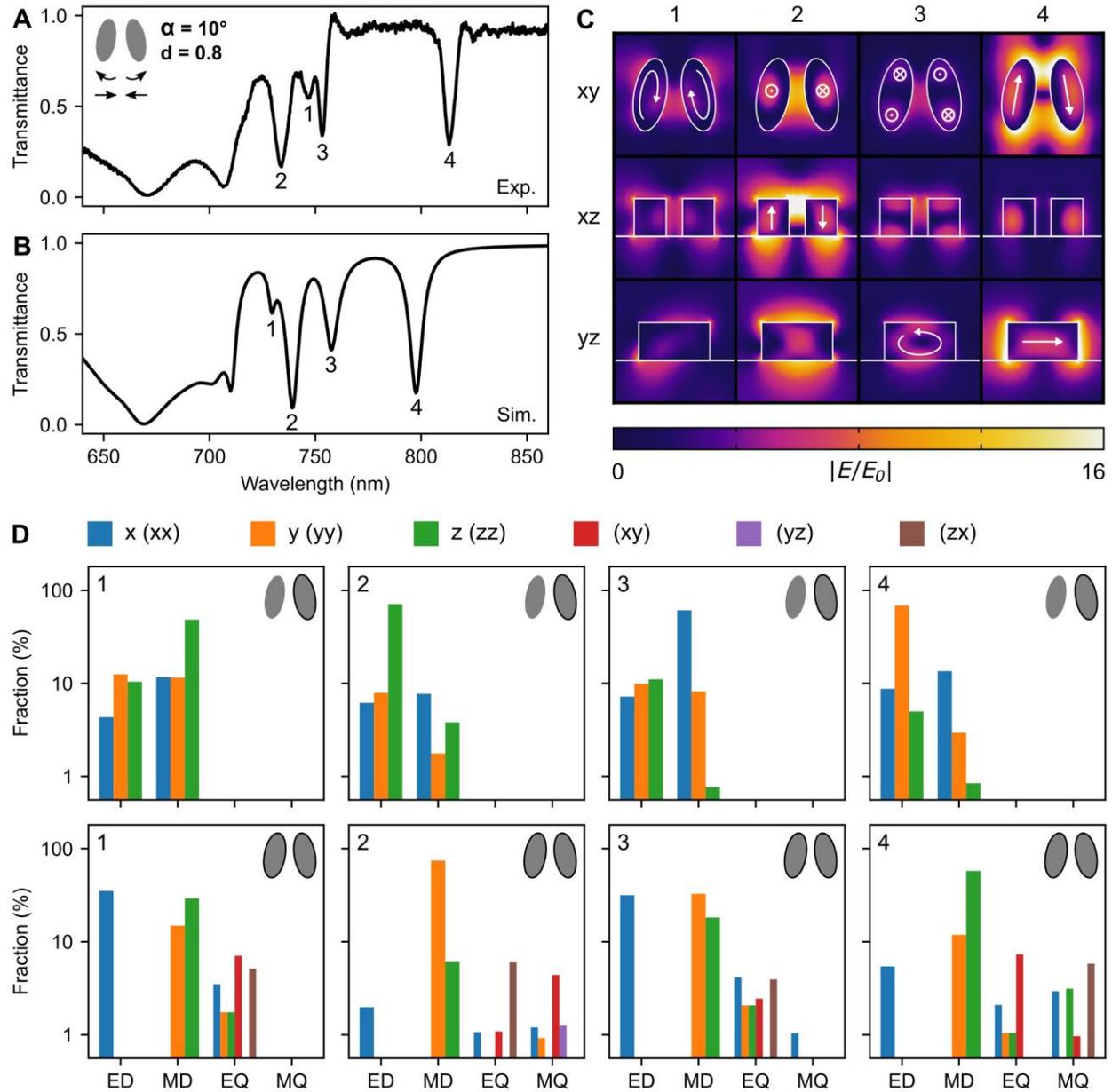

**Figure S2: Field distribution and multipolar decomposition of all resonances.** Experimental **(A)** and simulated **(B)** transmittance spectra for the d = 0.8, α = 10° metasurface design (Figure 3 of the manuscript), with the four states described in the main text labeled. **(C)** Electric field distributions, with each column corresponding to the resonance denoted at the top and each row to different cross-sections of the unit cell. **(D)** Electric dipole (ED), magnetic dipole (MD), electric quadrupole (EQ) and magnetic quadrupole (MQ) moments of the system with their cartesian fractions (x, y, z for dipoles and xx, yy, zz, xy, yz, zx for quadrupoles). Each column corresponds to the mode marked in the top-left corner of the bar-charts and each row to different integration volumes, marked as black outlines in the top-right corners.

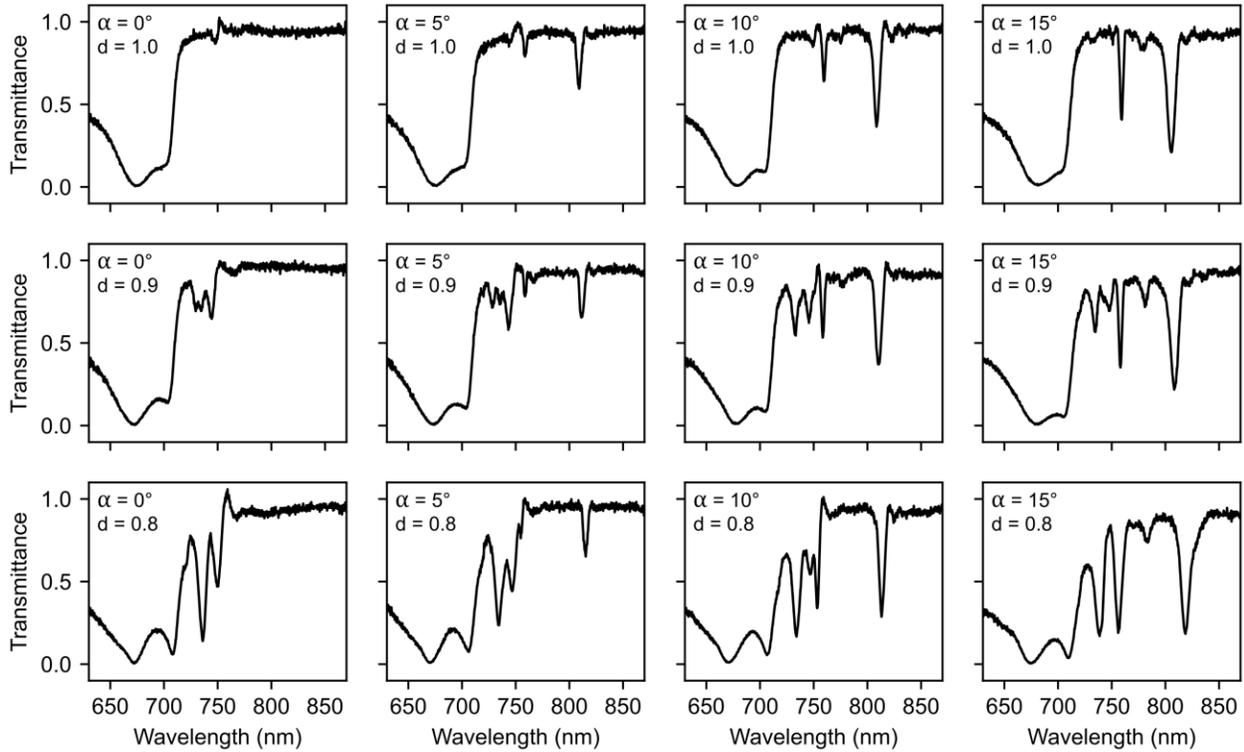

**Figure S3: Additional transmittance spectra for varying α and d.** Transmittance spectra of metasurfaces with α ranging from 0° to 15°, and d from 0.8 to 1. The top-left plot shows the fully symmetric metasurface, each successive column increases the angle, and each row below decreases the intra-cell meta-atom distance.

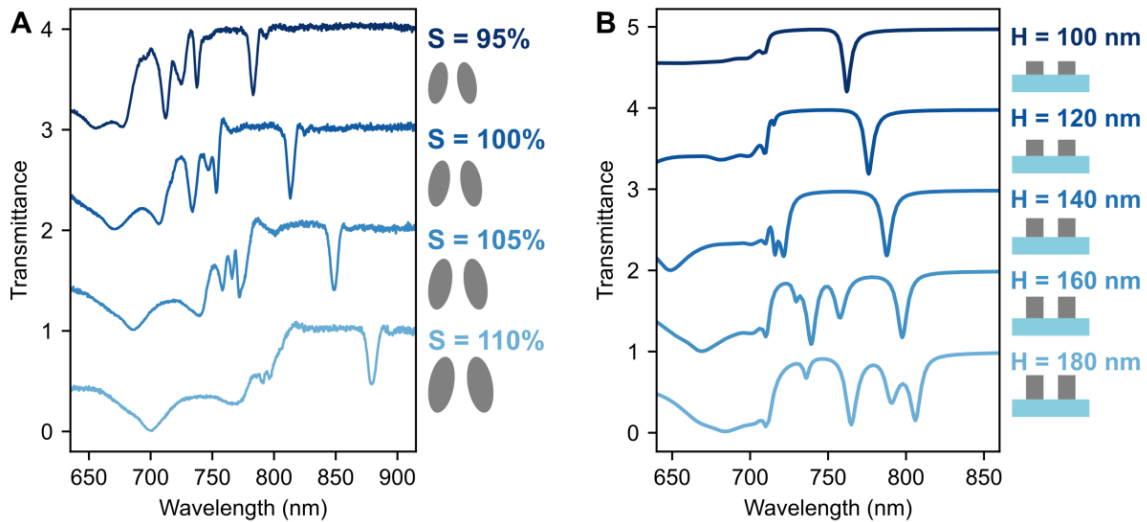

**Figure S4: Simulations for different lateral dimensions and height of the design. (A)** Transmittance spectra for different sizes of the unit cell. The height is kept fixed at 160 nm, and the ellipses diameter and periodicity are scaled (S) up or down; S = 100% corresponds to the nominal design. **(B)** Simulated transmittance for different metasurface heights (displayed on the right of the graph) while keeping the rest of the geometrical dimensions constant. It can be seen how a minimum height is required to support more than one qBIC resonance. The transmittance spectra are vertically displaced for clarity.

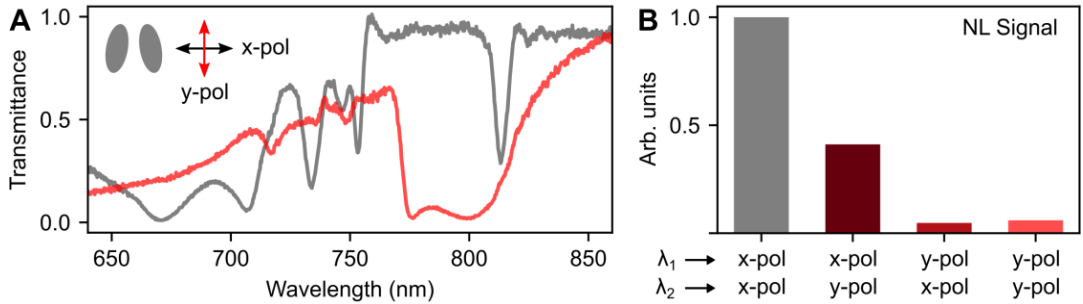

**Figure S5: Linear and nonlinear experiments with y-polarized light. (A)** Comparison of the transmission response with x-polarized and y-polarized light for the d = 0.8, α = 10° metasurface design. A scheme of the unit cell in the top-left corner shows the direction of the polarization with respect to the meta-atoms orientation. **(B)** Nonlinear signal at $\lambda_1$ = 741 nm and $\lambda_2$ = 816 nm (same as in Figure 4 of the manuscript) for different polarization conditions. The maximum signal is obtained when both pump lasers are x-polarized, and the minimum when the 'Pump 1' laser is y-polarized.

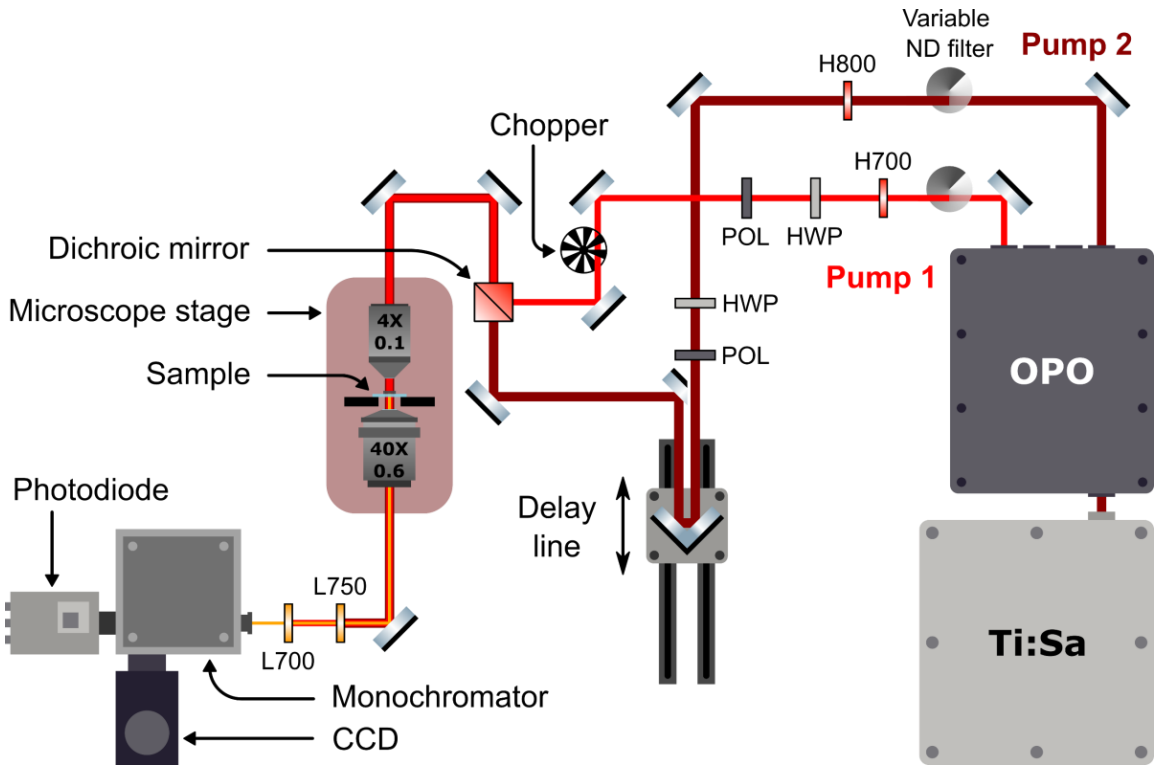

**Figure S6: Experimental setup for nonlinear measurements.** 'Pump 1' corresponds to the second harmonic of the OPO signal (500-800 nm) and 'Pump 2' to the Ti:Sapphire laser (700-1000 nm). Each arm before the sample has high-pass filters (H700 and H800), polarization optics (half-wave plates, HWP and linear polarizers, POL) and neutral density filters (ND). In this way the interference of the nonlinear signal with the tails of the lasers can be avoided and the polarization alignment and intensity on the sample is controlled. On the detection side two low-pass filters (L700 and L750) are used to measure the generated light at $\lambda_s$ ~ 640 - 690 nm. The delay line is used to temporally overlap both laser pulses at the sample position.

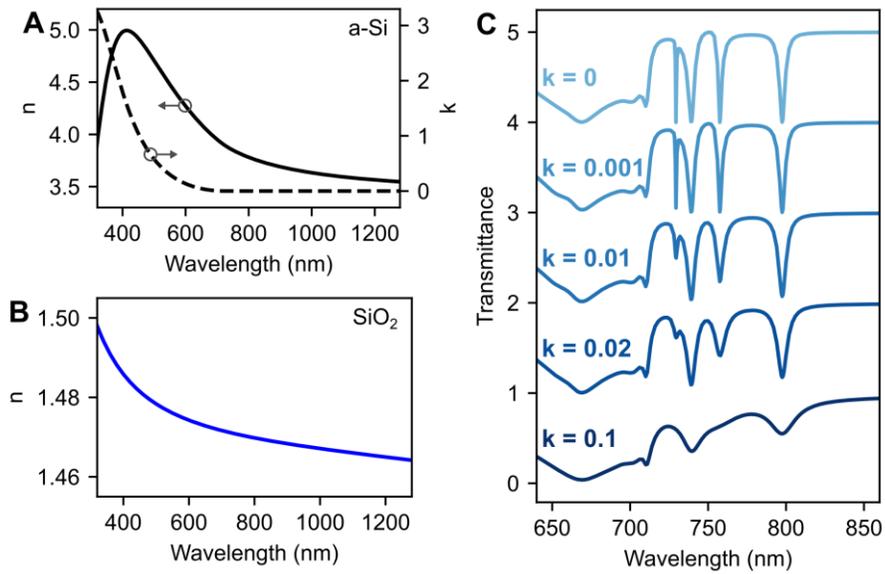

**Figure S7: Ellipsometry data for Si and SiO$_2$ and simulations changing k.** Refractive index (n) and extinction coefficient (k) for a-Si **(A)** and SiO$_2$ **(B)** obtained from ellipsometry measurements. **(C)** Simulations showing the transmittance spectra when increasing k, where a broadening of the resonances is observed. Although a-Si exhibits k < 0.001 above 700 nm, we considered a k value of 0.02 when comparing to experiments to better approximate other non-radiative losses, like surface roughness. The transmittance spectra are vertically displaced for clarity.